\documentclass{article}
\usepackage[utf8]{inputenc}
\usepackage{graphicx}
\usepackage{float}
\usepackage[normalem]{ulem}

\usepackage[a4paper,top=3cm,bottom=2cm,left=2cm,right=2cm,marginparwidth=1.75cm]{geometry}

\usepackage[citestyle=nature]{biblatex}

\usepackage{amsmath,amssymb}
\usepackage{multicol}
\usepackage[colorinlistoftodos]{todonotes}
\usepackage{gensymb}
\usepackage[normalem]{ulem}

\usepackage{appendix}
\usepackage{lineno}
\usepackage{authblk}
\author[1]{Ana H. Lobo*}
\author[1]{Andrew F. Thompson}
\author[2]{Steven D. Vance}
\author[2]{Saikiran Tharimena}

\affil[1]{California Institute of Technology, Pasadena, CA, USA}
\affil[2]{Jet Propulsion Laboratory, California Institute of Technology, Pasadena, CA 91109, USA }

\usepackage{setspace}

\usepackage{fancyhdr}
\title{A pole-to-equator ocean overturning circulation on Enceladus.}

\begin{document}

\maketitle

\vspace{10mm}
\begin{flushright}
\textbf{Under Review}\\
\vspace{3mm}
Comments can be sent to corresponding author: \\
\vspace{2mm}
Ana H. Lobo \\
California Institute of Technology\\
1200 E. California Blvd., M.C. 150-21\\
Pasadena, CA 91125\\
Email: lobo@caltech.edu\\
\end{flushright}

\clearpage
\doublespacing

Enceladus is believed to have a saltwater global ocean with a mean depth of at least 30~km \cite{Cadek2019,Hemingway2019}, heated from below at the ocean-core interface and cooled at the top \cite{Choblet2017}, where the ocean loses heat to the icy lithosphere above. This scenario suggests an important role for vertical convection to influence the interior properties and circulation of Enceladus' ocean.  Additionally, the ice shell that encompasses the ocean has dramatic meridional thickness variations that, in steady state, must be sustained against processes acting to remove these ice thickness anomalies. One mechanism that would maintain variations in the ice shell thickness involves spatially-separated regions of freezing and melting at the ocean-ice interface.  Here, we use an idealized, dynamical ocean model forced by an observationally-guided density forcing at the ocean-ice interface to argue that Enceladus' interior ocean should support a meridional overturning circulation.  This circulation establishes an interior density structure that is more complex than in studies that have focused only on convection, including a shallow freshwater lens in the polar regions. Spatially-separated sites of ice formation and melt enable Enceladus to sustain a significant vertical and horizontal stratification, which impacts interior heat transport, and is critical for understanding the relationship between a global ocean and the planetary energy budget. The presence of low salinity layers near the polar ocean-ice interface also influences whether samples measured from the plumes \cite{waite2017cassini} are representative of the global ocean. 

\section{Introduction}

The ocean of Enceladus is the best characterized and potentially most accessible of the many oceans in our solar system that have been studied to date  \cite{glein2019geochemistry}. The polar asymmetry in its surface geology and active plumes at its south pole are consistent with regional heating \cite{Choblet2017} that suggests chemical gradients on a global scale. \textit{Cassini} spacecraft measurements provided compelling evidence for a subsurface ocean, as inferred from the high heat output from the south pole \cite{Spencer2006}, which is global in nature based on thermodynamic arguments \cite{collins2007enceladus}, as well as the interpretation of the tiny moon's gravity and spin states \cite{mckinnon2015effect}. Particles and gases sampled from the south polar plume indicate organics in the ocean \cite{Postberg_2018}, and suggest a modestly high pH with chemical affinity supportive of methanogenesis \cite{waite2017cassini}. Models of tidal heating that sustains the ocean and the non-uniform ice thickness indicate flexure of a porous seafloor, and localized outflow of warm fluids in the southern regions \cite{Choblet2017}. This regional upwelling of warm fluid may also suggest strong, localized melting and associated mixing and fractionation at the ice--ocean boundary, where materials enter the stream of plume ejecta. 

Spatial heterogeneity in Enceladus' ice shell is a strong indicator that localized regions of freezing and melt at the ocean-ice interface modify the density of the subsurface ocean through heat exchange, freshwater fluxes, and brine rejection. 
This scenario is analogous to Earth's high latitude oceans where variations in surface buoyancy forcing due to interactions with sea ice and ice shelves have a leading order control on the large-scale circulation \cite{Marshall2012, Ferrari8753}.  Indeed, with some knowledge of interior mixing rates, the distribution of ocean surface density fluxes can constrain the structure and strength of the large-scale overturning circulation \cite{Marshall2003}.  We leverage this approach, commonly referred to as water mass transformation \cite{Groeskamp2019}, in our analysis of Enceladus' circulation.

A steady overturning circulation requires the transformation of waters between different density classes, implying some degree of stratification in the ocean interior.  Heat and freshwater exchange at the ocean-ice interface \cite{Soderlund2014} and through the generation of convective plumes \cite{Goodman2003} are key mechanisms for the production and destruction of water of different densities. If the production of water with distinct properties and densities is spatially separated, a circulation that connects these regions is required to close the overturning.   In particular, if the separation between regions of melt and freezing is comparable to the size of Enceladus’ ocean, then we predict that the ocean will support a large-scale, pole-to-equator circulation \cite{Zhu2017}.  Previous work has largely focused on thermal driving of ice covered oceans \cite{Soderlund2014} and its tendency to homogenize interior properties through mixing. Freshwater driven processes have received less attention, although previous studies have shown how this forcing can generate turbulent eddy transport \cite{Jansen2016} and a layered stratification \cite{Zhu2017}, while other studies have highlighted their potential importance \cite{Soderlund2019}.  We comment on the potential for freshwater formed at the ocean-ice interface to stratify the ocean interior below.

Linking the spatial distribution of freshwater fluxes, i.e. regions of melting and freezing, to observed variations in ice shell thickness provides insight into the circulation dynamics within Enceladus’ ocean. Constraints on the overturning circulation would improve estimates of oceanic heat, freshwater, and nutrient transport, helping to close the system’s energy budget and identifying how chemical gradients may be established and maintained. As we will show, the overturning circulation also affects the interpretation of measurements from Enceladus’ surface plumes, including prior constraints on the ocean’s composition and habitability \cite{waite2017cassini}. Here, we develop an idealized model to understand controls on the circulation within Enceladus’ ocean. Our approach is to explore 
the sensitivity of the ocean circulation's structure in response to changes in ocean characteristics, such as surface buoyancy forcing and interior mixing properties, including spatially-heterogeneous turbulent plumes.

\vspace{5mm}

\section{Methods}

The use of an idealized ocean model enables a broad assessment of plausible stratification and circulation regimes and their dependence on characteristics of Enceladus' ocean.   This model, summarized below and described in detail in \ref{extramethods}, enables us to (i) identify the primary dynamical balances that govern the circulation, (ii) explore possible global circulation configurations, and (iii) optimize parameter choices for more realistic but computationally-intensive general circulation models. This work will also aid in relating key parameters to observable properties from future missions.  An important assumption of the model is that the global ocean sustains some level of stratification.  While the density of each layer is prescribed, ocean stratification is determined from the model output based on the layer depths and thicknesses.  Salinity is assumed to provide the stratifying agent; in section \ref{TandS} we compare model densities to estimated temperature and salinity ranges for Enceladus.
  
Following similar terrestrial ocean applications \cite{Marshall2003, Thompson2019}, the model tracks the transport of water between a discrete number of fixed density classes.  Transfer between layers, or ``transformation'', occurs due to diabatic processes through two mechanisms: mixing in the interior or direct forcing at the ocean-ice interface.  We link the latter to spatial variations in the ice shell thickness. 

In steady state, the formation or melt of ice must occur where the ice shell is anomalously thick or thin, respectively, to sustain ice shell thickness gradients. This pattern of melting and freezing would oppose the smoothing tendency of the ice pump mechanism  \cite{lewis1986} that arises from changes in melt temperature as a function of pressure (\ref{extramethods}). The ice pump mechanism would be particularly strong for a thick shell in isostasy, but the exact rate depends on ocean circulation.  Here, we do not address the physical drivers behind the phase changes, but they could arise, for example, due to geothermal heat release or local upwelling could drive melting in localized regions with a thinner ice shell. Regardless of the cause, spatial variations in ice thickness imply significant forcing through density (freshwater) fluxes at the ocean-ice interface. 

Mixing in the ocean interior is assumed to occur primarily through rotationally-constrained geostrophic turbulence \cite{Speer2000, Jansen2016, Soderlund2014}. The source of this geostrophic turbulence that sustains the overturning circulation is itself influenced by the ice shell thickness and its associated pattern of freezing and melting.  Together these establish an ocean density that varies not only vertically, but also laterally.  Horizontal density gradients, represented in our model as tilting density layers, provide the source of potential energy from which ocean eddies form and grow through baroclinic instability \cite{Green70}.  These eddies both stir properties along density surfaces and give rise to an eddy volume transport (\ref{extramethods}).  This adiabatic motion, or motion within density classes, plays a critical role of delivering water between spatially-separated sites of water transformation.  

Interior mixing that governs the volume transport between density layers occurs through smaller-scale turbulence. In Earth's ocean, interior mixing arises due to the action of internal wave motions \cite{Munk1966}, whereas on Enceladus, the interior mixing may be significantly amplified by turbulent plumes \cite{Vance2018, Choblet2017} that may be spatially heterogeneous. At steady state, interior diabatic transport and adiabatic advection are balanced by transformation at the ocean-ice interface (fig.~\ref{fig:diagram}b).

The model evolves the interfaces between various density classes in two ways.  In the ocean interior, convergence of mass in a density class due to turbulent mixing, parameterized as a turbulent diffusivity $\kappa$, either changes the thickness of a layer through a vertical displacement $w_{int.}$ (\ref{extramethods} eq. \ref{eq_w}), or is balanced by adiabatic lateral transport.  The latter is incorporated using a well-tested closure \cite{GentMcWilliams90} that depends on the slope of density surface and an isopycnal eddy diffusivity $K_e$.  
The latitudinal position where density layers outcrop at the ocean-ice interface may also evolve in time, $v_{out.}$ (\ref{extramethods} eq. \ref{eq_v}), due to an imbalance between water supplied adiabatically to the interface and the transformation to a different density class. A solution for the overturning circulation can be determined by integrating these velocity equations in time until a steady state is achieved.

Model parameters enable us to assess how various physical properties of the Enceladus system influence the global circulation. The diapycnal diffusivity $\kappa$ describes the intensity of vertical mixing or the exchange across layers in the ocean interior. Increased tidal heat release at the ocean-core boundary could intensify turbulent mixing in the ocean interior and lead to a higher $\kappa$ overall or to regional increases in the case of ocean plumes. The isopycnal diffusivity $K_e$ captures the efficiency of baroclinic eddies in the ocean interior. The magnitude of $K_e$ could vary spatially, due to changes in the baroclinic deformation radius, for instance, but we apply a constant $K_e$ in the simulations described below for simplicity. 
The magnitude of the surface density (buoyancy) forcing $F_{b}$ represents the rate of ice formation/melt and its meridional distribution impacts the location where transformation takes place. Starting from a control experiment, parameters were systematically varied in the simulations described below.

\section{The dynamical balance of a pole-to-equator overturning}

\begin{figure}[h]
    \centering
    \includegraphics[width=0.9\textwidth]{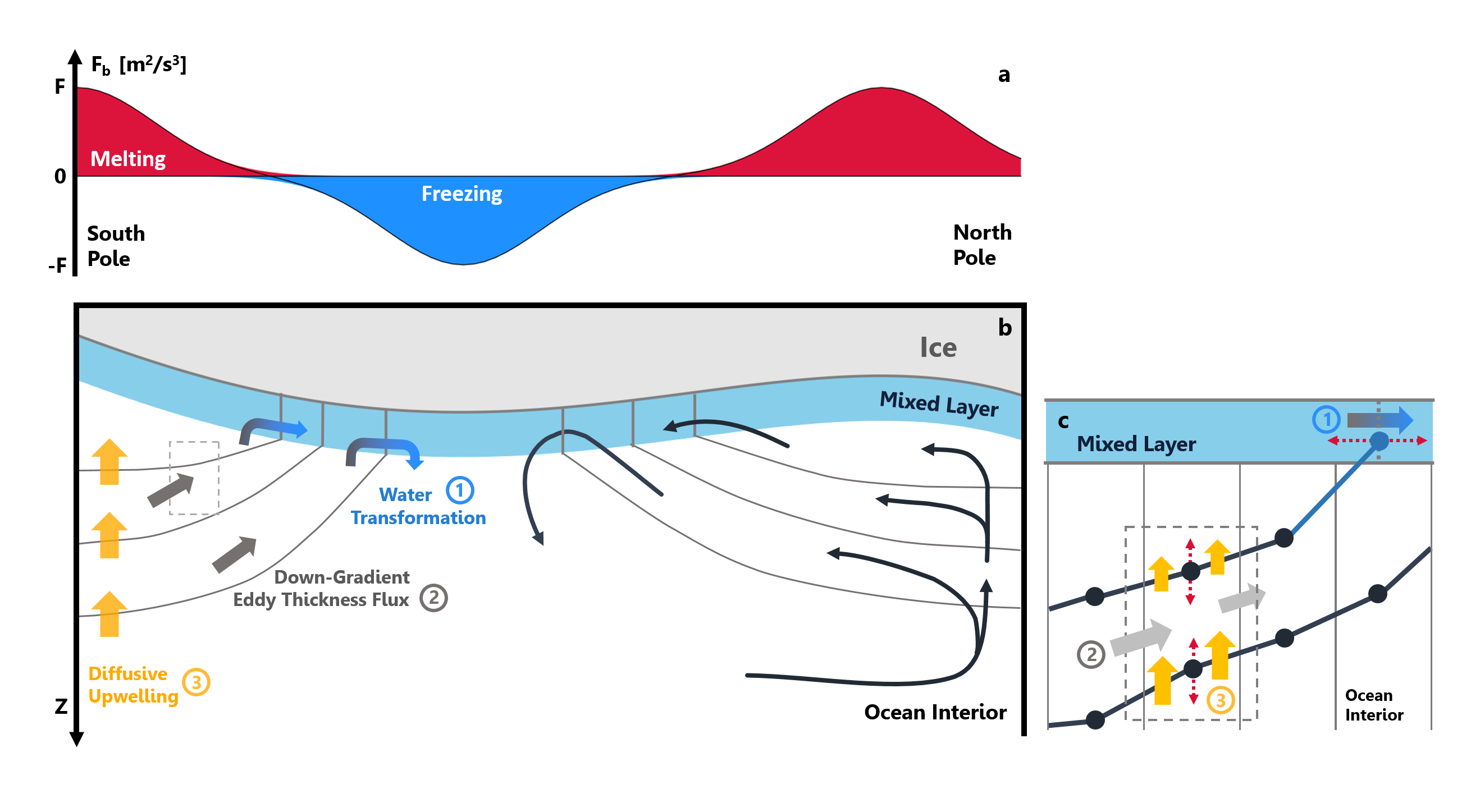}
    \caption{Diagram of Enceladus-like ocean and circulation.
    (a) Inferred surface buoyancy forcing ($F_{b}$, $m^2/s^3$), based on measured meridional ice thickness gradients shown in panel (b).  (b) Schematic of the ocean interior divided into layers of different densities (grey lines). The right hand side provides a schematic view of the total overturning circulation, while the left hand side illustrates the physical processes that support this circulation. 
    The inset in panel (c) shows how the physical processes are represented within a model grid. The dark blue dots indicate the depth of each layer that evolves with time. The position where the interface outcrops at the ocean-ice boundary (light blue dot) also evolves in time, independent of the grid. This figure is not drawn to scale.  Simulations described below focus only on the Southern Hemisphere, with a simplified ice gradient (see fig.~\ref{fig:layers_control}) and surface buoyancy forcing (see \ref{extramethods} fig.~\ref{fig:model_setup})}
   \label{fig:diagram}
\end{figure}

To provide intuition for the dynamics that influence the ocean stratification and circulation for a given surface forcing, we first describe an equilibrated state for a set of control parameters. For this simulation, the ocean-ice interface is 10~km deeper at the equator than at the pole and the buoyancy flux distribution $F_{b}(y)$ has a maximum amplitude of $10^{-8}$ m$^2$~s$^{-3}$ (fig.~\ref{fig:diagram}a). The magnitude of interior mixing ($\kappa \approx 5 \times 10^{-4}$ m$^2$~s$^{-1}$) is laterally uniform throughout the ocean except near the poles where we prescribe enhanced diapycnal mixing to represent convective plumes. Values for the control parameters are provided in \ref{extramethods}.

The density structure of the control simulation (fig.~\ref{fig:layers_control}) illustrates that an overturning circulation is sustained in a relatively shallow layer below the ice shell. The stratification only penetrates $\sim$ 1~km below the ocean-ice interface, but varies from the pole to the lower latitudes where these density surfaces intersect the ocean-ice boundary layer.  Density layers are thicker in the polar region, leading to a deeper, but weaker stratification.  Closer to the equator the density layers shoal and the stratification intensifies. This stratification also generates a density-layer thickness gradient (fig.~\ref{fig:control_h_and_phis}a) that supports an equatorward adiabatic circulation that draws water towards the ocean-ice boundary layer and converges mass into the region of sea ice formation at lower latitudes. The water entering the ocean-ice boundary layer is transformed into denser water by brine rejection near the equator and is subducted back into the deep interior. The flow in the deep ocean (below layer 5), not explicitly resolved in this model, would have a poleward flow to conserve mass. This overturning circulation is balanced by both interior diabatic mixing that lightens water masses and the high-latitude melting that sustains the lightest (freshest) layer.

Enhanced polar mixing, represented by an increase in $\kappa$, strengthens the rising, adiabatic branch of the circulation at the poles. This also produces an increase in the total volume transport or magnitude of the overturning circulation (fig.~\ref{fig:control_h_and_phis}b), illustrating the tight connection between adiabatic and diabatic motions. The overturning circulation is present even with a uniform interior mixing (fig.~\ref{fig:phase_diagram}g), as might be the case if there were no polar plumes or if plumes were evenly distributed.

We next consider the sensitivity of the stratification and circulation to key parameters in the model, and seek to relate these parameters to observable quantities. To characterize the ocean's overall structural changes in response to varying parameters, we examine two properties:  (i) the ``outcrop'' position, or the meridional location where each layer interface intersects the ocean-ice boundary, and (ii) the penetration depth of the stratification, equivalent to the depth of each layer interface at the pole.  

The layer outcrop positions are strongly constrained by the distribution of the surface buoyancy flux.  A poleward shift in the ice formation region ($\phi_b$) confines the low density layers to the higher latitudes (fig.~\ref{fig:phase_diagram}a). This more compact configuration, because of the increased layer slopes, results in a stronger adiabatic transport (fig.~\ref{fig:phase_diagram}b). Meanwhile, a decrease in the ice formation rate, equivalently the buoyancy flux, or an increase in eddy activity, $K_e$ (fig.~\ref{fig:phase_diagram}c,e) causes the density layers to spread equatorward, equilibrating in regions where the mixed layer transformation is stronger (closer to $\phi_b$). For high $K_e$, the increased lateral flux results in a stronger, but shallower, overturning in the upper ocean (fig.~\ref{fig:phase_diagram}f). 

In contrast to the parameters described above, the stratification depth is more sensitive to the magnitude of the vertical mixing $\kappa$ (fig.~\ref{fig:phase_diagram}g). An increase in $\kappa$ enhances the diabatic exchange between density layers, increasing the lateral layer thickness gradients. This results in both a stronger lateral transport (fig.~\ref{fig:phase_diagram}h) and a deeper overturning circulation. To balance the stronger overturning, layers outcrop in regions where water mass transformation is more intense. Therefore, while strong vertical mixing (high $\kappa$) and strong baroclinic eddy activity (high $K_e$) have opposite effects on layer depth, both lead to a broader and stronger overturning circulation. 

\vspace{10mm}

\section{Discussion \& Conclusion}

The idealized numerical model supports a few simple but powerful statements about the ocean circulation in Enceladus.  First, if ice formation occurs predominantly at mid to high latitudes, an overturning circulation can be geographically confined.  As regions of ice formation extend to lower latitudes, a pole-to-equator overturning emerges. Furthermore, a stratified ocean with a freshwater lens in the polar regions is sustainable across a broad range parameter space and the equatorward extent of this lens is bounded by regions of net ice growth. These results imply a direct link between the surface buoyancy forcing at the ocean-ice interface and the interior ocean stratification and circulation. We expect that future measurements of the ice shell will allow for better estimates of this buoyancy flux through higher resolution ice-shell thickness distributions, but also by constraining the dynamics and composition of the ice shell.  In future simulations, the connection between ice shell thickness and buoyancy flux should be evaluated in a model where the ocean circulation is coupled to an evolving ice shell. 

Our results have focused on the circulation, but oceanic flow also involves the transport of heat and other tracers. While recent tidal heat dissipation studies support that the ocean may be primarily heated from below at the poles \cite{Choblet2017}, these patterns do not agree with the more complex ocean-ice heat flux pattern that is believed to be required for the stability of the ice shell \cite{Cadek2019}. A significant meridional overturning could carry a portion of the high-latitude geothermal heat flux to lower latitudes via a lateral, likely eddy-driven, circulation. Similarly, freshwater exchange at the ocean-ice interface on Enceladus can establish regions of upwelling away from the poles directed towards sites of brine rejection, in a manner that is in qualitative agreement with the ocean heat flux required by the ice shell \cite{Cadek2019}. Thus, a pole-to-equator overturning can reconcile these two seemingly contradictory boundary conditions and conflicting views of Enceladus.

The circulation will also impact how nutrients from the ocean-mantle boundary are distributed. If nutrients are transported upward primarily by ocean plumes, we would expect detrainment to release these nutrients in the ocean interior. In Earth's ocean, it has been shown that adiabatic mixing, or stirring along isopycnals, can be an important pathway for nutrient delivery to the surface and modulate productivity there \cite{Uchidaetal20}.  

Thus, the distribution of freshwater fluxes at the ocean-ice interface, by setting global patterns of upwelling, could influence nutrient fluxes and provide insight into which regions of an icy world have resources for life to potentially flourish.

Thus far, \textit{Cassini} measurements of Enceladus' polar surface plumes have provided the best insight into Enceladus' ocean composition. Based on studies of freshwater driven processes \cite{Zhu2017}, we suggest that low density layers tend to form where the ice is thinnest, as is the case for Enceladus' south pole. Given the expected temperature range for Enceladus \cite{Vance2018}, the thermal expansion coefficient for ocean water would be very small ($5.788 \times 10^{-5}$ K $^{-1}$ at 20MPa, see \ref{TandS}). Therefore, salinity differences are likely the dominant source of density variations. If we assume that the ocean has a mean salinity equal to or larger than that of our deepest layers, the ``fresh water'' lens feeding the surface plumes would have a salinity at least 2~g/kg lower than the ocean mean (\ref{TandS}).

The dynamics we describe here for Enceladus are typical of geophysical fluids influenced by stratification and rotation and are therefore likely to occur in other ocean worlds. Most notably, \textit{Cassini} observations  indicate Titan's ice shell is 50-100~km thick, with possible spatial variability in thickness exceeding 10~km \cite{choukroun2012titans,Durante_2019}, 
large enough to alter the global density structure of the ocean. In larger ocean worlds like Titan, lateral density gradients may also arise from interactions with the ocean bottom, where a second lithosphere of high-pressure ice resides \cite{Vance2018}. The effects of such variations on the compositional and thermal structure of the ocean might be measured by their influence on the ocean's seismic and electrical properties \cite{Vance2018}.  The \textit{Dragonfly} mission, planned to study Titan in further detail starting in 2034 \cite{turtle2017dragonfly}, will carry a geophysical package that may yield clues to the properties of the ocean. In the Jupiter system, the \textit{Europa Clipper} and \textit{JUICE} missions, which will study Jupiter's moons Europa and Ganymede, respectively, in the coming decade \cite{grasset2013JUICE,buffington2017evolution}, will constrain any variations in lithospheric thickness while also probing the electrical properties of the subsurface oceans. Our results highlight the critical need for more detailed studies into the global circulation structure of ocean worlds.

\section*{Acknowledgements}
A portion of this research was carried out at the Jet Propulsion Laboratory, California Institute of Technology, under a contract with the National Aeronautics and Space Administration (80NM0018D0004). This work was partially supported by JPL's strategic research and technology program, and by the Icy Worlds node of NASA’s Astrobiology Institute (13-13NAI7 2-0024). \copyright 2020. All rights reserved. AFT was supported by the David and Lucile Packard Foundation.

\printbibliography

\newpage

\section*{Figures}

\clearpage

\begin{figure}[h]
    \centering  
    \includegraphics[width=0.8\textwidth]{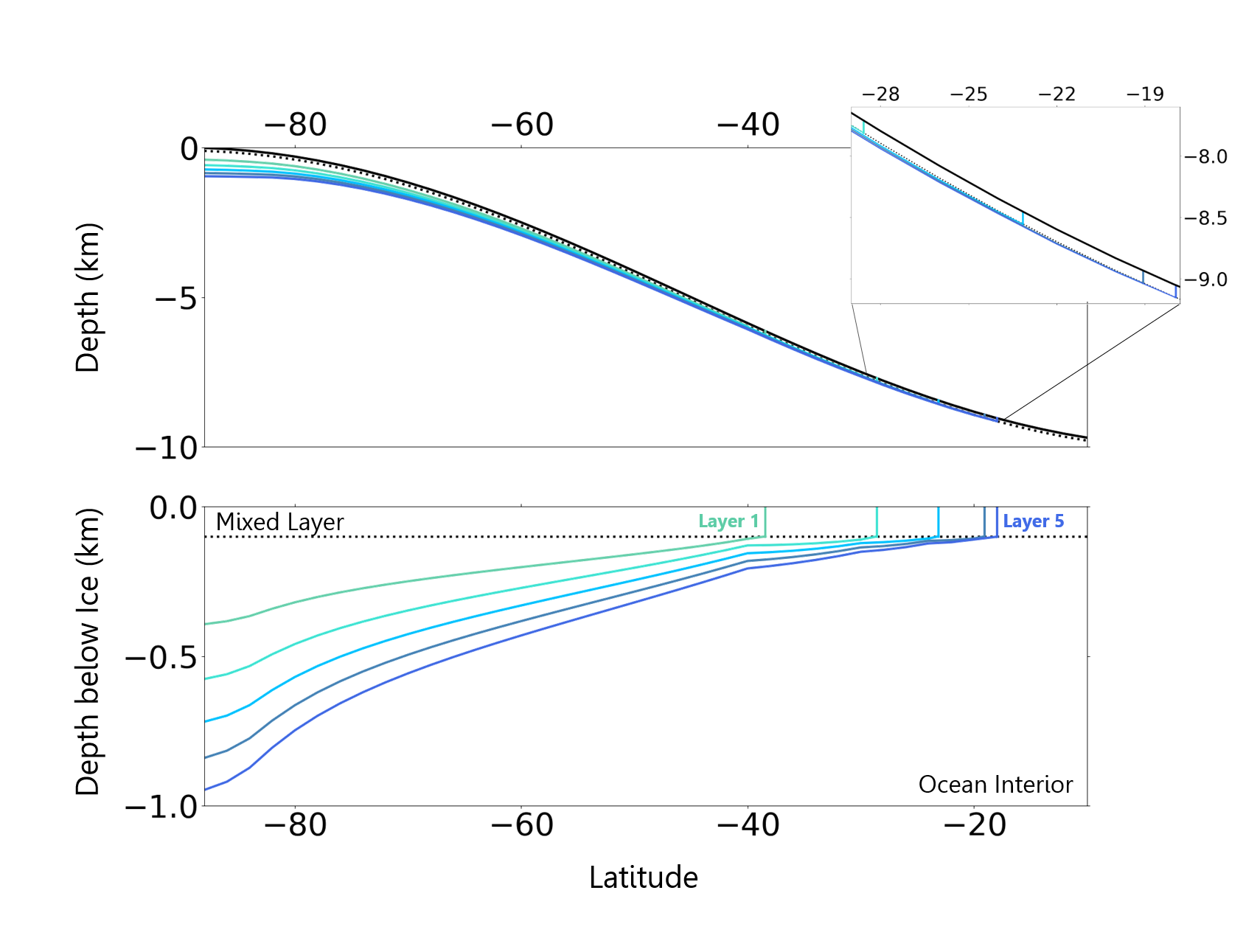}
    \caption{Steady state distribution of density layers in the control run. The top panel shows the position of the layer interfaces with respect to the sloped ocean-ice interface. The bottom panel shows the same layers, plotted as depth below the ocean-ice interface to more easily visualize gradients in the layer thickness. }
    \label{fig:layers_control}
\end{figure}

\clearpage

\begin{figure}[h]
    \centering
    \includegraphics[width=0.8\textwidth]{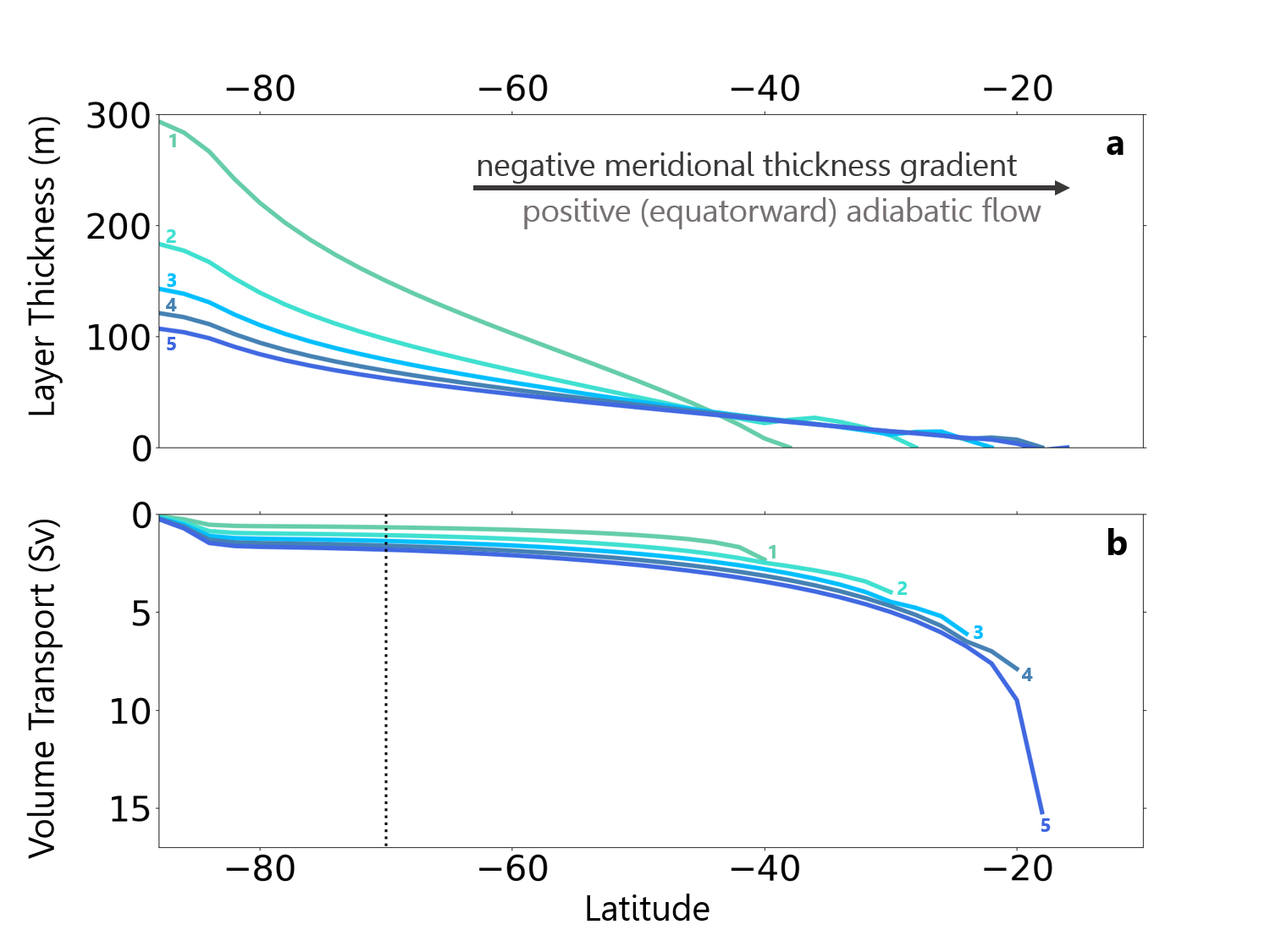}
    \caption{Characteristics of the control simulations. (a) Thickness of each of the five density layers (m) as a function of latitude.  The reduction in layer thickness moving towards the equator supports an equatorward eddy transport within each layer.  (b) Volume transport (1 Sv = 10$^6$ m$^3$ s$^{-1}$) across density surfaces, integrated from the pole northward, such that  $\Psi_{Diabatic}(\phi) = \int_{-90}^{\phi} w \, 2 \pi R \cos({\phi^\prime}) \, \mathrm{d}\phi^\prime$, where $\phi^\prime$ is an integration variable. Positive values indicate a volume flux from denser to lighter layers.  At steady state, the diabatic transport across a given interface, integrated between the pole and latitude $\phi$, must be balanced by the total adiabatic lateral transport above that interface, such that $\Psi_{Diabatic} = \Psi_{Adiabatic} = K_e (s - s_{ice}) \, 2 \pi \cos{\phi}$. The black dashed line at 70\degree \space latitude indicates the location where lateral fluxes are calculated for fig.~\ref{fig:phase_diagram}. The $y$-axis has been inverted so that the deepest density layer (5) is shown at the bottom and the lightest density (1) is at the top. In both panels, the layer numbers and colors are the same as in fig. \ref{fig:layers_control}. }
    \label{fig:control_h_and_phis}
\end{figure}

\clearpage

\begin{figure}
    \centering
    \includegraphics[width=1\textwidth]{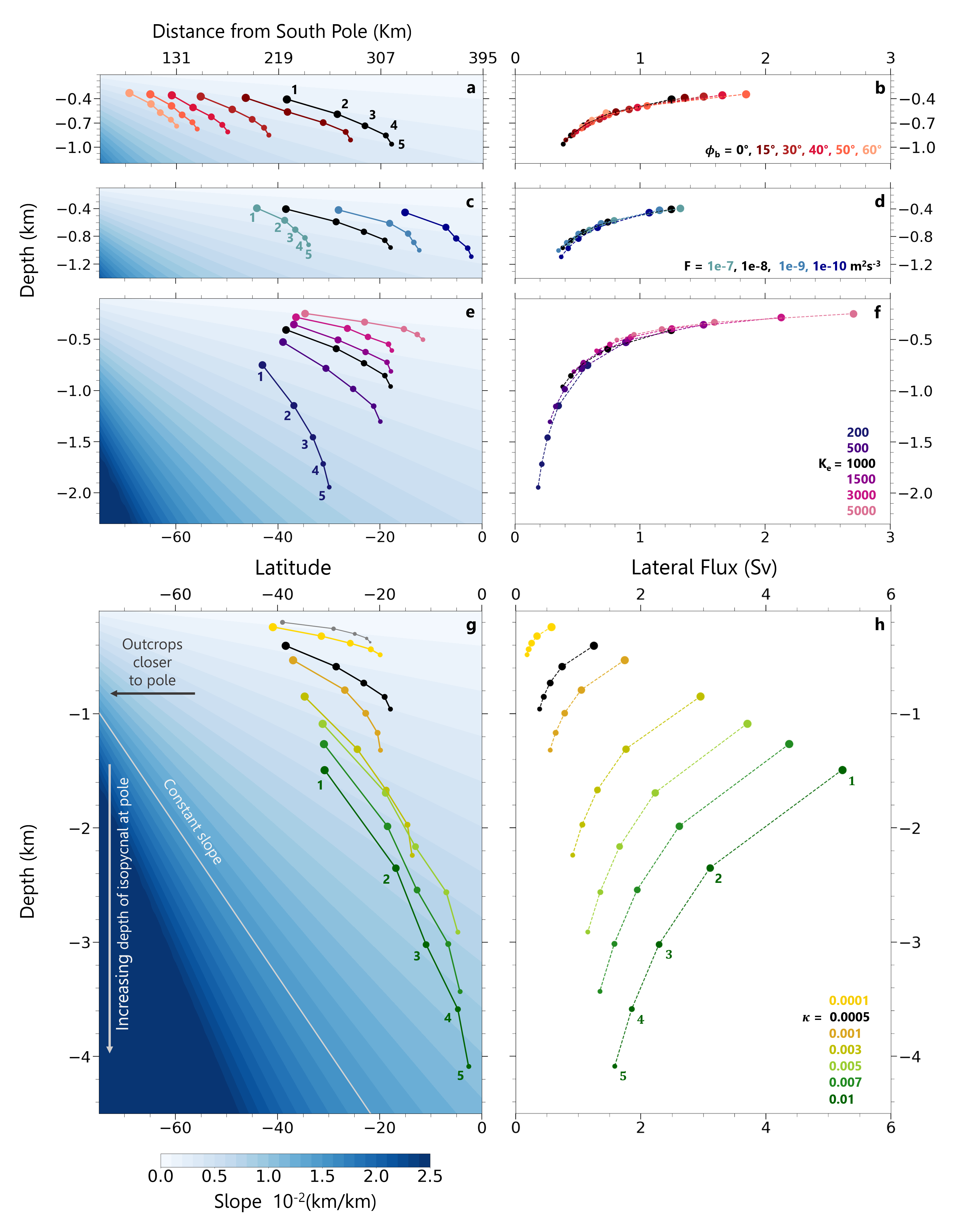}
    \caption{Phase diagram depicting isopycnal slope, for varied model parameters. From top to bottom, we show results from varying $\phi_{brine}$ ($^{\circ}$ latitude),  $F$ (m$^2$ s$^{-3}$), $K_e$ (m$^2$ s$^{-1}$), $\kappa$ (m$^2$ s$^{-1}$). For the panels on the left, the vertical axis shows the depth (km) of each layer interface at the south pole and the horizontal axis shows the latitude of the interface outcrop location. The five outcrops of each model are plotted with markers of the same color. The background colors indicate lines of constant isopycnal slope. The right panels show the lateral volume flux (1 Sv = 10$^6$ m$^3$ s$^{-1}$)  within each layer at 70\degree \space latitude. In panel (g), the model results shown in grey are for a control-like simulation where polar amplification of diapcynal mixing $\kappa$ has been removed. }
    \label{fig:phase_diagram}
\end{figure}

\clearpage

\renewcommand\thesection{S.\arabic{section}} 
\setcounter{section}{0} 
\renewcommand\theequation{s\arabic{equation}} 
\setcounter{equation}{0} 

\section*{Supplementary Information}

\section{Expanded Methods}\label{extramethods}

In this section, we describe the idealized overturning model in greater detail. We begin with the conservation equation for buoyancy. Buoyancy is linearly related to density $\rho$ through the relationship,
\begin{equation}
b = g \left( \frac{\rho_0-\rho}{\rho_0} \right) , 
\end{equation}
where $\rho_0$ is a reference density.  The use of buoyancy emphasizes that differences in density constrain the overturning circulation as opposed to the absolute density. Conservation of buoyancy is given by:

\singlespacing
 \begin{equation}
    \frac{\partial b}{\partial t} = -\mathbf{u}\cdot\nabla b - \nabla\cdot F_b \approx -\underbrace{v b_y \, - \: w b_z}_{\text{advection}} + \underbrace{\frac{\partial}{\partial z}\left(\kappa\frac{\partial b}{\partial z}\right)}_{\parbox{38pt}{\scriptsize\centering interior mixing}} - \underbrace{\frac{\partial F_b}{\partial z}}_{\parbox{38pt}{\scriptsize\centering buoyancy flux at sfc.}}.
    \label{fred}
\end{equation}
\doublespacing

In steady state, $\partial b/\partial t = 0$ such that both adiabatic and diabatic advection in the ocean interior are balanced by transformation at the ocean-ice interface. The flow, represented by $v$ and $w$, is the total ocean circulation comprised of both mean and eddy components, for example $v = \overline{v} + v^\prime$, more commonly referred to as the residual circulation in the context of terrestrial water-mass transformation models \cite{Groeskamp2019}.

Assuming the upper part of Enceladus' ocean is zonally unbounded, the mean meridional velocity is weak and the meridional transport is dominated by eddy fluxes, e.g. $\overline{v^\prime b^\prime}$ and $\overline{w^\prime b^\prime}$. The mean meridional velocity, $\overline{v}$, vanishes because, assuming a small Rossby number, there is no zonally-averaged zonal pressure gradient to support this flow. We parameterize the eddy transport using a well-tested closure from the oceanographic literature \cite{GentMcWilliams90}, in which the adiabatic component of the advection arises from the relationship $\Psi = K_e s L_x(y)$, such that $v = -\partial \Psi/\partial z$ and $w = \partial \Psi/\partial y$.  Here $s$ is the slope of the interface separating density layers, $L_x(y)$ is the zonal length of each latitude circle, $\Psi$ is a streamfunction quantifying a volume transport, and $K_e$ is an isopycnal eddy diffusivity (m$^2$ s$^{-1}$).  The eddy advection is assumed to act along density surfaces so that $v$ and $w$ combine to be parallel to the layer interfaces. We set $K_e$ to an Earth-like value (1000~m$^2$~s$^{-1}$) for the control simulation, and we test the sensitivity of the circulation to changes in $K_e$. 

An important assumption in these simulations is that $K_e$ is uniform throughout the ocean.  Mixing length theory is often employed to argue that the eddy diffusivity scales as  $K_e\sim U\ell$, which depends on both the strength of the eddy velocities $U$ and the size of the eddy $\ell$ \cite{Vallis06}. We can approximate the eddy mixing length as the Rossby deformation radius $R = NH\pi^{-1}f^{-1}$. Here, $f$ is the Coriolis parameter ($f = 2 \Omega \sin\phi$) and $N^2 = \partial b / \partial z$. If we use $N^2 \approx \Delta b/ H$ , where $H$ is the total ocean depth (30~km), we would have $R = 10$~km in the midlatitudes. If we instead calculate $N$ based on our model outputs and use our deepest layer as a measure of the pycnocline such that we can set $H$ to that layer's depth, this results in $R = 3$~km (for 45\degree \space latitude). Given that the total distance between the equator and pole on Enceladus is 395~km, this deformation radius provides reasonable support for a scale separation between the eddy and domain sizes and therefore justification for using an eddy diffusivity.  As the magnitude is poorly constrained, we have varied this parameter in our simulations. While $K_e$ might be expected to have meridional structure (likely increasing from pole to equator), we expect that only those simulations that have an overturning with a larger meridional extent, \textit{e.g.} low values of $\phi_b$ and $F_b$ and high values of $\kappa$, would be sensitive to these variations. 

We also include the diabatic transport across density surfaces in the ocean interior, which occurs due to mixing at scales smaller than the mesoscale. This turbulent mixing is parameterized by a small-scale turbulent eddy diffusivity $\kappa$, which supports a vertical advection-diffusion balance in the ocean interior, $wb_z = \left(\kappa b_{z}\right)_z$, where subscripts indicate partial derivatives.  In the isopycnal model the vertical velocity can be approximated by 
    \begin{equation}
    w(y) \approx \frac{\kappa}{\Delta z_n},
\end{equation}
where $\Delta z_n$ is the thickness of the layer above the interface in question \cite{Thompson2019}.  This approximation is valid if vertical variations in $\kappa$ are small relative to the buoyancy gradient ($\kappa_z b_z \ll \kappa b_{zz}$). Note that diabatic streamfunction is defined such that $
    \Psi_{diff}(y) = \int_{-90}^{y} w(y')L_x(y')\,\mathrm{d}y'$.

The last term on the right hand side of eq.~\ref{fred} accounts for the buoyancy forcing at the ocean-ice interface. The buoyancy forcing arises from both differential heat uptake (changing the water temperature) and phase changes (changing salinity); we assume the latter to be the dominant cause of buoyancy differences on Enceladus. Phase changes (melting--refreezing of ice and aqueous exsolution--dissolution) must occur to compensate for processes that smooth out the meridional ice thickness gradients. 

Over sufficiently long time scales, thick ice sheets deform plastically leading to ice transport along the thickness gradient \cite{Goodman2003}. However, this flow is proportional to the gravitational force, which means it should be weak on a small moon such as Enceladus ($g = 0.113$ m s$^{-1}$), but could be significant on Europa ($g = 1.315$ m s$^{-1}$) and Titan ($g =1.352$ m s$^{-1}$). An ice pump mechanism \cite{lewis1986} could also be active, particularly for a thick shell in isostasy. The pump mechanism is fueled by the change in melt temperature as a function of pressure. Under the deeper regions of the ice shelf (at higher pressure) the melt point is lower, and the water at the interface is colder than that at shallower interfaces. If this cold water is displaced (by tides or other mechanisms), it will move up to the shallower regions where it will serve as a heat sink and promote freezing. This process would tend to reduce thickness gradients in the ice shell at a rate proportional to the temperature gradient along the base of the ice shelf. The rate is proportional to the interface depth variations, such that we would expect tens of meters of melt per Enceladus year, but the specific rate also depends on the ocean circulation itself. Thus, sustained ice shell thickness gradients require ice to form in thicker regions and melt in thinner regions and Enceladus' ice thickness pattern implies the existence of significant buoyancy forcing at the ocean-ice interface. 

Water within density layers that intersect this mixed layer, or outcrop, interact with the buoyancy forcing and are subject to water mass transformation. In the model, we prescribe a buoyancy flux $F_{b}$ at the interface in each hemisphere of the form:
\begin{equation}
    F_{b}(y) = F \bigg\{ exp \Big[- \left( \frac{|y - \phi_m| }{\sigma} \right) ^2 \Big]
    - exp \Big[- \left( \frac{|y - \phi_b| }{\sigma} \right) ^2 \Big] \bigg\},
\end{equation}
where $F$ is the forcing magnitude (in m$^2$~s$^{-3}$). Regions of melt are centered at the poles ($|\phi_m |= 90$\degree \space = 396~km), and the forcing decays exponentially over a length scale of $\sigma = 90$~km. Regions of ice growth and brine rejection are centered at the equator ($\phi_b = 0$\degree \space = 0~km) for the control run, but are varied for other simulations. Transformation in the ocean surface boundary layer is quantified by the relationship \cite{Marshall2003}:
\begin{equation} \label{eq:phi_res}
\Psi_{F} = F_{b}(\partial b / \partial y)^{-1}, 
\end{equation}
where the last term is the meridional buoyancy gradient within the ocean-ice boundary layer region.

With parameterizations for the terms in eq.~\ref{fred} in hand, a solution for the ocean circulation can be determined by integrating in time to a steady state. In the mixed layer, where we have a buoyancy forcing and $b_z$ is by definition negligible, we can divide both sides of eq.~\ref{fred} by the meridional buoyancy gradient at the interface ($b_y|_{sfc}$), to get: 

\begin{equation}\label{eq_v}
v_{out.}  = \frac{\partial y}{\partial t} = \frac{1}{h} \left( K_e s - \frac{F_{b}}{b_y|_{sfc}}\right),
\end{equation}
which describes the time evolution of the outcrop location.
Meanwhile, in the ocean interior, where there is no influence from the buoyancy forcing, we can divide by $b_z$ to arrive at:
\begin{equation}\label{eq_w}
w_{int.} =  \frac{\partial z}{\partial t} \approx \left( 
{K_e\frac{\partial s}{\partial y}} + \frac{\kappa}{\Delta z} \right),
\end{equation}
where $\partial_t z$ is the vertical velocity of the layer interface at a given location. Eq. \ref{eq_w} can also be derived directly from conservation of mass in the ocean interior. 

Through the assumption of zonal symmetry, this model does not explicitly include zonal flows.  However, it does not preclude the existence of a zonal circulation that coexists with the meridional circulation, similar to Earth's atmosphere.  The presence of zonal jets \cite{Soderlund2014} could generate a frictional stress at the interface that establishes a time-mean component of the overturning circulation, an Ekman layer, near the ocean-ice boundary \cite{Marshall2003}.

We summarize these processes as two key balances: (i) the flow from the ocean interior towards the ocean-ice interface must balance the water mass transformation in the mixed layer. Until this balance is met, any net convergence in the mixed layer will induce density layer outcrop displacement. (ii) Any net volume transport, through vertical mixing, into a density layer in the ocean interior must be transported along the layer and eventually flow into the mixed layer. If the net down-gradient flow at any point is not balanced by the net diabatic mixing, the convergence within a layer induces a change in layer depth.

Our control run, shown in fig.~\ref{fig:layers_control} and \ref{fig:control_h_and_phis}, utilized the parameters shown in fig.~\ref{fig:model_setup} and listed in Table~\ref{tab:table1}. The latter also itemizes the full range of parameters explored in fig.~\ref{fig:phase_diagram}, plus the range of meridional ice thickness gradients ($\Delta_{ice}$) that were studied. The $\Delta_{ice}$ results were not included because changes to the ocean structure and flow were small.
Note that we are only modeling a few discreet layers, which limits our resolution of the circulation patterns. Our parameters were selected to facilitate comparison over a wide range of simulations, but follow-up work interested in the deeper ocean circulation would either require simulations varying $\Delta b$, or a more intense diabatic mixing.

\renewcommand\thefigure{s\arabic{figure}} 
\setcounter{figure}{0} 
\renewcommand\thetable{s\arabic{table}}

\clearpage

\begin{table}[p]
    \centering
    \begin{tabular}{|c|c|c|c|}
    \hline
    Model Parameter & Control value & Min & Max  \\
    \hline
    $\Delta b$ &  $ 4 \times 10^{-5}$ m/s$^2$ & - & - \\
    \hline
    $K_e$ - Isopycnal diffusivity & $10^{3}$ m$^2/s$ & 200 m$^2/s$ & 5000 m$^2/s$ \\
    \hline

    $\kappa$ - Diapycnal diffusivity & $5 \times 10^{-4}$ m$^2/s$ & 10$^{-4}$  m$^2/s$ & $10^{-2}$  m$^2/s$   \\
    \hline
    $\Delta$ice & ${10}$ km & $1$ km & $20$ km  \\
    \hline
    $\phi_{b}$ & 0\degree  & 0\degree & 60\degree\\ 
    \hline
    $F$ & $10^{-8}$ m$^2$/s$^3$ & 10$^{-10}$ m$^2$/s$^3$ & 10$^{-7}$  m$^2$/s$^3$ \\
    \hline
    \end{tabular}
    
    \caption{Simulation parameters for control run and the range of values tested for parameters that were explored.}
    \label{tab:table1}
\end{table}

\clearpage

\begin{figure}[h]
    \centering
    \includegraphics[width=0.8\textwidth]{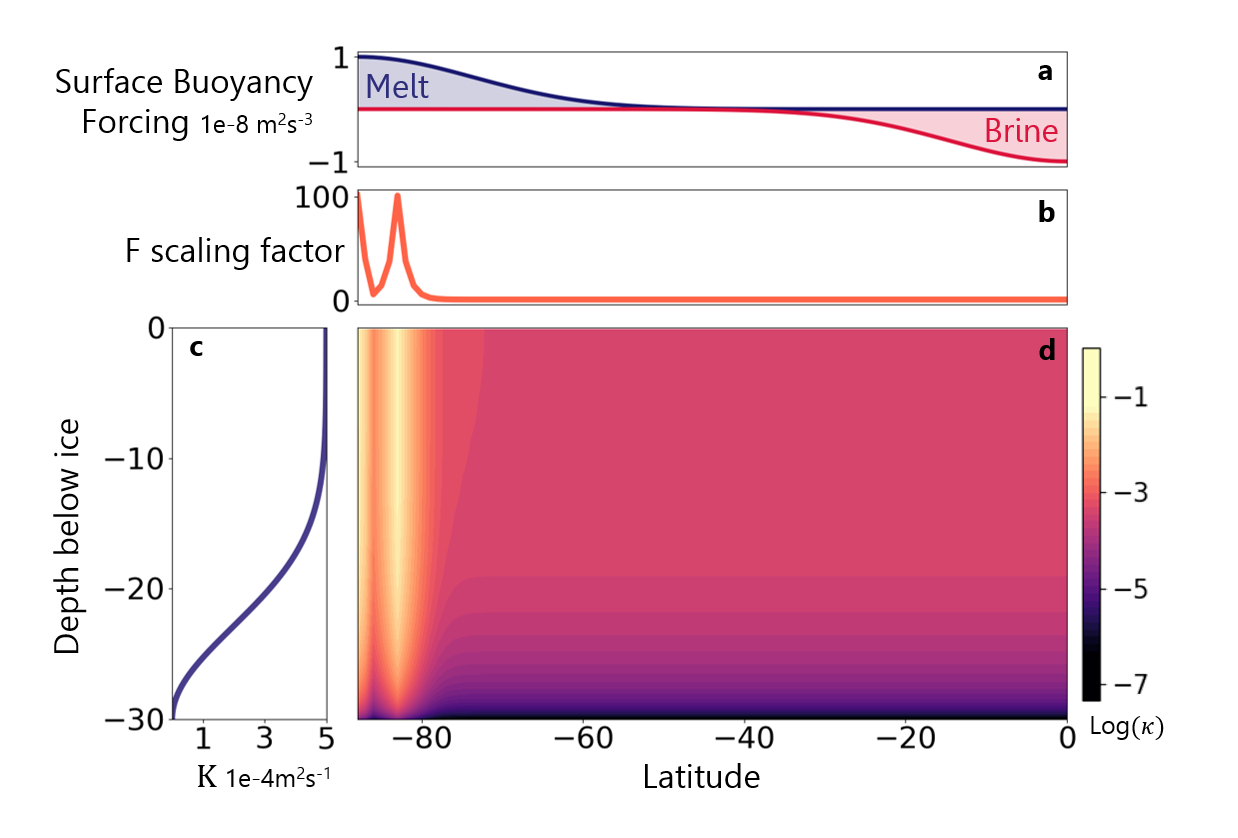}
    \caption{Depiction of key model parameters for control run values. (a) The prescribed surface buoyancy forcing ($F_{b}$) as a function of latitude for the southern hemisphere. The remaining panels show the components of the diapycnal diffusivity ($\kappa$). We include a scaling factor that varies with latitude (b), which is used in simulations that have enhanced polar mixing (plumes). $\kappa$ also has a vertical dependence (shown in c). The spatial distribution of $\kappa$ (logarithmic scale, e.g. $-1=10^{-1}$) is illustrated in the contour plot (d). }
    \label{fig:model_setup}
\end{figure}

\clearpage

\section{Ocean Temperature and Salinity}\label{TandS}

\begin{figure}[h]
    \centering
    \includegraphics[width=0.7\textwidth]{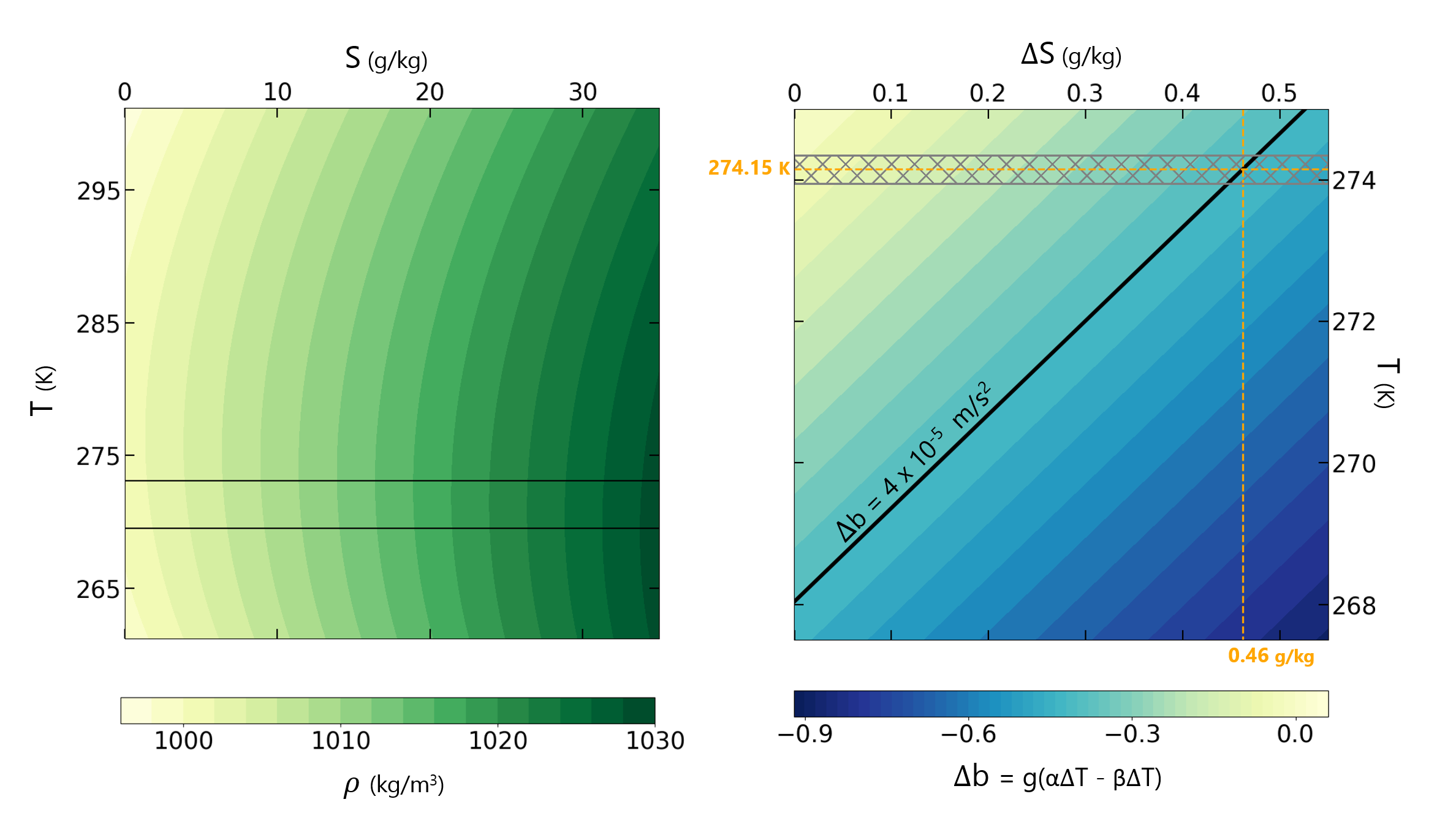}
    \caption{(Left) Predicted density for various temperatures and salinities at 20MPa. The black lines highlight the temperature range relevant for Enceladus \cite{Vance2018}. (Right) Equivalent change in temperature and salinity to produce the buoyancy difference ($\Delta_b$) between layers in the model, calculated at 20MPa, with a mean temperature of 274.15K. The black line highlights the model's $\Delta b$. For comparison, the grey hatching shows the expected variations in ocean temperature if the composition is kept constant (obtained using TEOS-10).  }
    \label{fig:s1}
\end{figure}

Our control model parameters (Table \ref{tab:table1}) were initially selected based on Earth-like values, although the value of $\Delta b$ was intentionally chosen to be small. Given that Enceladus' ocean is nearly an order of magnitude deeper than Earth's, our parameters were optimized for a weak stratification. However, we find that the isopycnals concentrate in the upper portion of the ocean in the control simulation, producing stronger stratification near the interface and weaker stratification in the deep ocean. We do not account for bottom boundary processes that could affect the deeper ocean density structure. 

The buoyancy gradients in the upper ocean could occur due to variations in temperature, salinity or a combination of both, and our model does not distinguish between these scenarios. However, for the predicted pressure and temperature range in Enceladus' ocean (fig.~\ref{fig:s1}, left), the equations of state predict that the density variations are almost entirely controlled by salinity. 

Using the linearized equation of state, a thermal expansion coefficient $\alpha = 5.788 \times 10^{-5}$~K$^{-1}$ and a saline contraction coefficient $\beta = 7.662 \times 10^{-4}$~kg~g$^{-1}$ (obtained using TEOS-10 for 20MPa), we can verify the changes required to produce a buoyancy difference of $\Delta b  = 4 \times 10^{-5}$~m~s$^{-2}$ (the difference between two sequential layers). These coefficients were obtained using a base temperature of 274~K, a degree higher than estimated in the literature \cite{Vance2018}, to remove negative thermal expansivity  arising from the lower assumed salinity of 12~ppt consistent with \textit{Cassini} measurements \cite{glein2019geochemistry}. These effects are probably relevant near the ice shell \cite{melosh2004temperature}, but are beyond the scope of this work. Under these conditions, if the ocean had constant salinity, a 6~K temperature difference would be required between layers (30~K overall). This is more than three times the temperature difference required for Earth-like conditions (1.7K). Whereas, for an isothermal ocean, a 0.46~g~kg$^{-1}$ salinity change between layers could account for the difference in buoyancy (fig.~\ref{fig:s1}, right).

Assuming that temperature variations in the ocean are limited to less than 4~K \cite{Vance2018}, the buoyancy differences modeled in this work and the ensuing circulation can be attributed almost entirely to salinity differences. The total variation in salinity between our lowest density layer and our highest density layer is estimated to be roughly 2~g~kg$^{-1}$. Further constraints could be obtained in future work through the use of additional model layers.

\clearpage

\end{document}